\documentclass[aps, twocolumn, superscriptaddress, floatfix]{revtex4-2}

\usepackage{graphicx}
\usepackage{dcolumn}
\usepackage{bm}
\usepackage{subcaption}
\usepackage{geometry}
\usepackage{array}
\usepackage{wrapfig,lipsum}

\newcolumntype{P}[1]{>{\centering\arraybackslash}p{#1}}

\begin{document}

\title{Pressure-dependent magnetotransport measurement in Kagome metal Yb\textsubscript{0.5}Co\textsubscript{3}Ge\textsubscript{3}}.

\author{Zhiyuan Cheng}
\email{Cheng@Physics.LeidenUniv.nl}
\affiliation{Leiden Institute of Physics, Universiteit Leiden, Niels Bohrweg 2, 2333 CA Leiden, the Netherlands}

\author{Yaojia Wang}
\affiliation{Kavli Institute of Nanoscience, Delft University of Technology, Lorentzweg 1, 2628 CJ Delft, the Netherlands}

\author{Heng Wu}
\affiliation{Kavli Institute of Nanoscience, Delft University of Technology, Lorentzweg 1, 2628 CJ Delft, the Netherlands}

\author{Mazhar N. Ali}
\affiliation{Kavli Institute of Nanoscience, Delft University of Technology, Lorentzweg 1, 2628 CJ Delft, the Netherlands}

\author{Julia Y. Chan}
\affiliation{Department of Chemistry and Biochemistry, Baylor University, Waco, Texas 76798, United State}

\author{Semonti Bhattacharyya}
\email{bhattacharyya@physics.leidenuniv.nl}
\affiliation{Leiden Institute of Physics, Universiteit Leiden, Niels Bohrweg 2, 2333 CA Leiden, the Netherlands}

\begin{abstract}
    Kagome materials are known to be an ideal platform that hosts a plethora of interesting phases such as topological states, electronic correlation, and magnetism, owing to their unique band structure and geometry. We report magnetotransport measurement in Kagome metal Yb\textsubscript{0.5}Co\textsubscript{3}Ge\textsubscript{3} as a function of pressure. Below $\sim25$ K the temperature dependence of resistance shows an upturn that is accompanied by a strong negative magnetoresistance, which could be attributed to Kondo effect. Upon pressurization above 1 GPa the resistance shows a reduction as a function of temperature below 4 K, along with a further enhanced negative magnetoresistance. This might indicate an onset of a pressure-induced Kondo coherence effect.
    
\end{abstract}

\maketitle

\section{Introduction}
Kagome lattice, which derives its name from a Japanese weaving technique, can be described as hexagons surrounded by corner-sharing equilateral triangles. Kagome materials, i.e. materials that host 2D Kagome lattice as part of their crystal structures have garnered a lot of attention in condensed matter physics, owing to their unique electronic and magnetic properties~\cite{Yin_Kagome_rev}. These unique properties stem from the Kagome crystal structure that leads to geometric frustration, as well as, band structure with rich features such as Dirac point, van Hove singularity, and flat band. Hence, Kagome materials have become a popular hunting ground for novel condensed matter phenomena such as topological physics, correlated electrons, charge density waves, and superconductivity -- all of which can co-exist~\cite{Wang_Kagome_rev}. It is important to understand the interrelation of these various phases to gain an understanding of the underlying physics of these materials. In this regard, perturbations such as charge doping or external pressure, that can enhance a phase and suppress another in a tunable manner, has proven to be really useful~\cite{zhao2021cascade_Kagome_dope_1,li2023unidirectional_Kagome_dope_2,li2022discovery_Kagome_pressure_1,zheng2022emergent_Kagome_pressure_2}. 
\begin{center}
\begin{figure*}[htbp]

    \centering
    \includegraphics[scale=0.85]{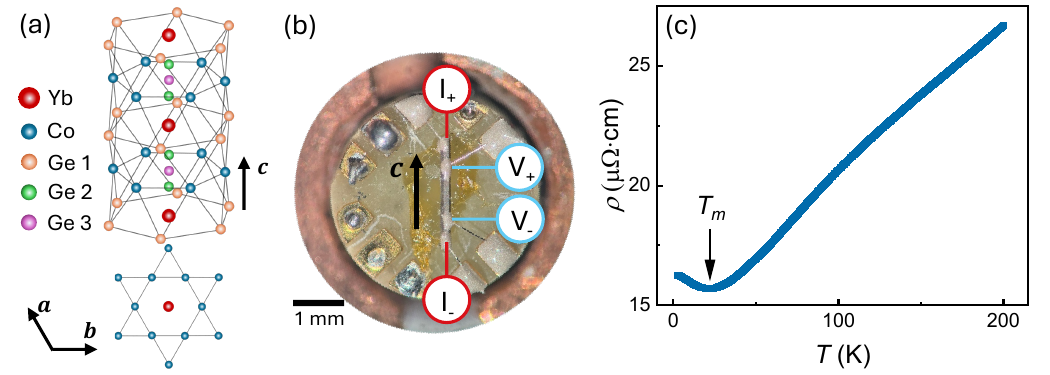}
    \caption{(a) Schematics of Yb\textsubscript{0.5}Co\textsubscript{3}Ge\textsubscript{3} crystal structure. (b) An image of the sample mounted on the PCB board of the pressure cell. The sample is contacted in 4-probe geometry. The current is passed from I+ to I-, and voltage is measured between V+ and V-. The direction of the current is aligned to the c-axis of the crystal shown in (a). (c) Temperature ($T$)-dependent resistivity ($\rho$). The black arrow indicates the onset of an upturn of resistivity with decreasing temperature.}
    \label{fig1}
   
\end{figure*}
\end{center}

In this work, we present pressure-dependent electrical transport measurements in Yb\textsubscript{0.5}Co\textsubscript{3}Ge\textsubscript{3} a recently-synthesized Kagome material~\cite{JChan}. The Kagome layer in this material hosts Co atom, with Yb inserted in between the consecutive Kagome plane. This unique geometry of Yb\textsubscript{0.5}Co\textsubscript{3}Ge\textsubscript{3}, along with the presence of both \textit{d}-shell and \textit{f}-shell electrons present us with an unique opportunity to investigate pressure-dependence of correlation, magnetism and their interaction in Kagome systems~\cite{tan2024three}.
Previous measurements in these materials have revealed the presence of antiferromagnetic exchange interaction between the Yb\textsuperscript{3+} moments without any long-range order~\cite{JChan}. Additionally, the low-temperature electrical transport measurements have revealed the presence of charge density wave (CDW) at $T_{CDW}=95$ K, and a resistivity upturn at $~18$ K that is associated with a magnetic transition~\cite{Wang_Kagome_Yb}. In this work, we further investigate the mechanism of this resistivity upturn with pressure as a tuning knob.
\section{Materials and Methods}
The Yb\textsubscript{0.5}Co\textsubscript{3}Ge\textsubscript{3} was prepared with the same method as reported previously~\cite{JChan,Wang_Kagome_Yb}. The samples were connected with silver epoxy for electrical transport measurements. The cryogenic measurements were performed with a physical properties measurement system (PPMS-Dynacool). The electrical transport measurements were performed with a 4-probe delta method measurement technique using a model 6221 sourcemeter and a model 2182A nanovoltmeter. The pressure-dependent measurements were performed in a PPMS-compatible pressure cell that is customized in-house for mounting samples perpendicular to the magnetic field on a PCB board.

\section{Results and Discussion}


Figure \ref{fig1} shows preliminary electrical transport measurement of  Yb\textsubscript{0.5}Co\textsubscript{3}Ge\textsubscript{3} crystal~\cite{JChan}.  Figure \ref{fig1}a) shows the crystal structure of 
Yb\textsubscript{0.5}Co\textsubscript{3}Ge\textsubscript{3}~\cite{JChan}. The Kagome planes are made of cobalt, interspersed by planes made of Ge atoms, and the Yb atoms are positioned in the Ge plane above and below the center of the Co hexagon. The crystal hosts two types of magnetic species -- Co with its \textit{d}-orbitals and Yb with its \textit{f}-orbitals. However, according to the susceptibility measurements performed in these samples, only Yb\textsuperscript{3+} magnetic moments contribute to the magnetic characteristics of the sample. Figure \ref{fig1}b) shows an image of the device mounted in the pressure cell. The pressure cell has been customized for mounting samples on a PCB board. The ring of Cu surrounding the PCB is the pressure feed-through holder. The single crystals grow in a needle shape with the \textit{c}-axis of the crystal aligned along the axis of the needle. The average size of the crystals is 3 mm. We have performed resistivity measurements in a 4-probe geometry, where current is applied in the \textit{c}-axis between the probes I\textsubscript{+} and I\textsubscript{-}, and the voltage is measured between V\textsubscript{+} and V\textsubscript{-}. Figure \ref{fig1}c) shows the temperature ($T$)-dependence of resistivity ($\rho$). Down to $\sim25$ K, the sample shows a metallic behavior as previously reported. 

    \begin{figure*}[t]
    \centering
    \includegraphics[width=\textwidth]{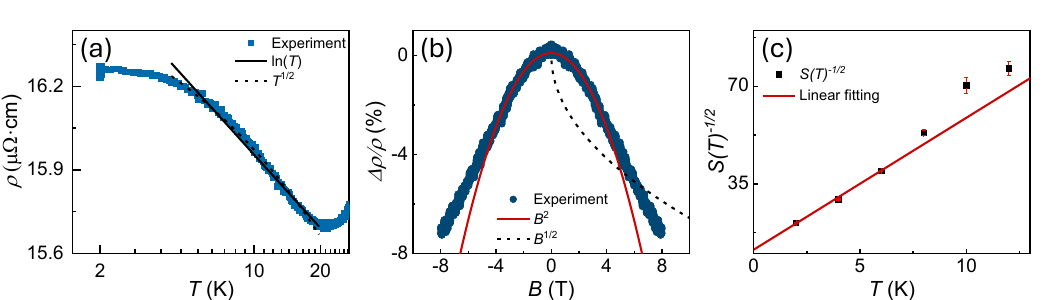}
    \caption{(a) A semilogarithmic plot of temperature ($T$)-dependent resistivity ($\rho$) at temperatures below 27 K. The solid black line and the dashed black line show logarithmic fit and $T^{1/2}$ fit respectively. (b) Magnetoresistance (MR) measurement ($\Delta\rho/\rho$ - $B$) at 2 K. The red continuous line and the black dashed line show $B^2$-dependence and $B^{1/2}$ dependence respectively. (c) $T$-dependence of $S(T)$, where $S(T)$ is obtained by fitting MR at different temperatures with equation~\ref{yosida}. The red line shows the linear fitting of the data according to equation~\ref{ST}.}
    \label{fig2}
\end{figure*}

Below $T_m\sim25$~K, $\rho-T$ shows a clear upturn (indicated by the black arrow) of resistivity with decreasing temperature in agreement with the previous report from Wang et al~\cite{Wang_Kagome_Yb}. This upturn has been attributed to a magnetic transition, However, the microscopic mechanism of this magnetic transition is not understood. In the subsequent part of this paper, we will investigate the origin of this upturn and study its pressure dependence.

It is important to note that, Wang et al.~\cite{Wang_Kagome_Yb} also observed a charge density wave (CDW) transition at $T_{CDW} = 95$ K in Yb\textsubscript{0.5}Co\textsubscript{3}Ge\textsubscript{3}. Our data does not demonstrate any clear signature of CDW. However, this is not surprising because in general, the signature of CDW is not always very clear in the resistivity measurement. Hall measurement is known to be a better probe of CDW. However, we were unable to contact these samples by Hall probes, due to needle-like shape of the samples.

Figure \ref{fig2} further elucidates the origin of this low-temperature upturn. Figure \ref{fig2}a) shows the $\rho$ - $T$ below $T_m$ in semilogarithmic scale. In a metallic diffusive system, such an up-turn at low temperature is usually attributed to either quantum interference effects such as (1) electron-electron interaction (EEI), (2) weak localization (WL)~\cite{PLee&TRamakrishnan} or magnetic scattering effect such as (3) Kondo effect~\cite{YKatayama_PRL_1966,YKatayama_PR_1967}. It is well-known that in a 3D case, both EEI and WL lead to $T^{1/2}$-dependence of resistance~\cite{PLee&TRamakrishnan,mani1991weak_3D_WL}, whereas Kondo effect results in $ln(T)$-dependence. Given the narrow temperature range of our data, it fits well with both types of $T$-dependence without any significant difference. Hence, based on this analysis we could not exclude any of the above-mentioned mechanisms.

We performed magnetoresistance (MR) measurements to further investigate the origin of this upturn. Figure \ref{fig2}b) shows magnetoresistance data obtained at 2~K. Magneto-transport measurement shows a strong negative MR below $T_m$ and MR is proportional to $B^2$ up to around 4 T. As EEI contributes to positive MR~\cite{PLee&TRamakrishnan}, we can exclude EEI as the mechanism of resistance-upturn below $T_{m}$. In contrast, both the WL and the Kondo effect are supposed to result in negative MR. WL correction to the MR follows a $B^2$ dependence at low fields, i.e. $B\leq B_{l}$, and $B^{1/2}$ dependence at higher fields~\cite{mani1991weak_3D_WL}. Here, $B_l$ is given by the equation,
 \begin{equation}\label{WL equation}
    B_l=\hbar/4eD\tau_l
\end{equation}
where $D$ is the diffusivity of the charge carriers and $\tau$ is the phase coherence time. An overestimation of $B_l$ is around 100 mT ($D~10$ cm\textsuperscript{2}$\cdot$s, $\tau~0.01$ 0.01 ns). However, it is clear in Figure \ref{fig2}b) that MR shows $B^2$-dependence up to magnetic fields ($B\sim4$~T) that are orders of magnitude higher than our estimated $B_l$, whereas the MR at high field does not follow $B^{1/2}$-dependence at all. Thus, having excluded both WL and EEI as the mechanism of temperature and magnetic-field dependence of resistance, we are left with the Kondo effect as the only possible mechanism.
\begin{center}

\begin{figure*}[htbp]
    \centering
    \includegraphics[width = \textwidth]{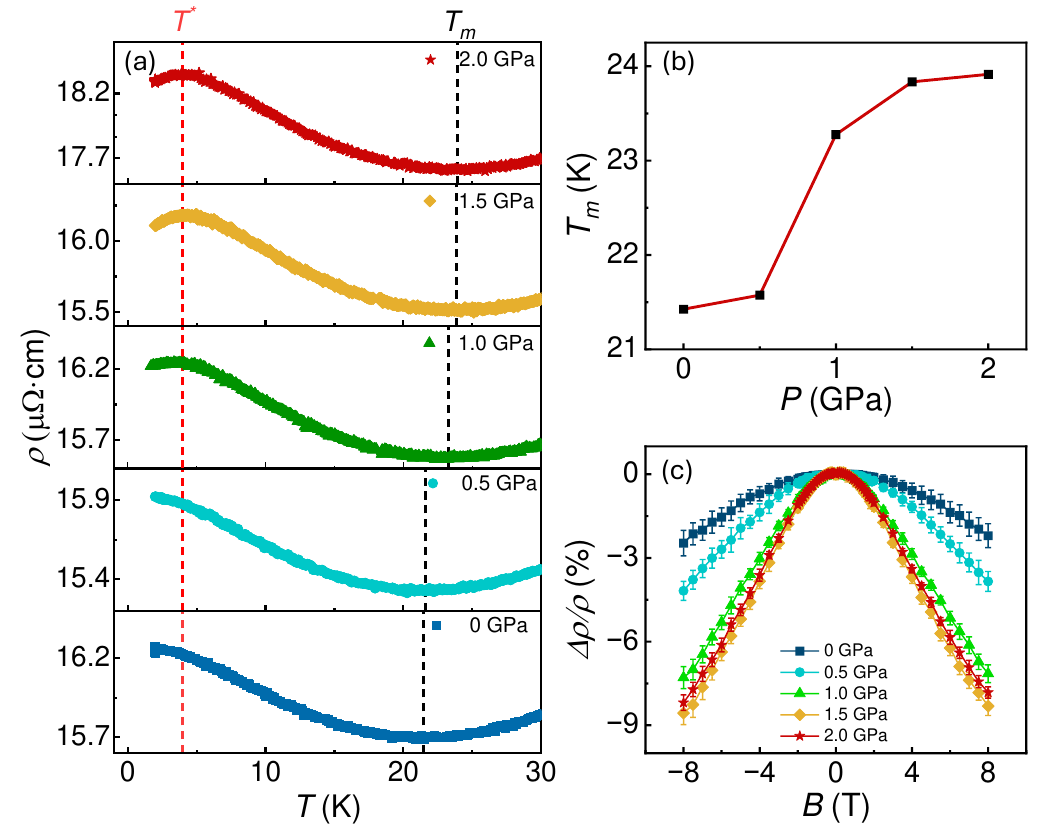}
    \caption{Pressure-dependent transport measurements performed on Yb\textsubscript{0.5}Co\textsubscript{3}Ge\textsubscript{3}. (a) Temperature ($T$)-dependence of resistivity ($\rho$) measured at pressures of 0 GPa (blue), 0.5 GPa (cyan), 1 GPa (green), 1.5 GPa (yellow) and 2 GPa (red). The black dashed line and the red dashed line respectively indicate minima and maxima in the ($\rho$-$T$) plot. (c) Magnetoresistance ($\Delta\rho/\rho$-$B$) measured at 2 K under pressures of 0, 0.5, 1, 1.5 and 2 GPa.}
    \label{fig3}
\end{figure*}
\end{center}


According to Yosida's theory of localized spins, negative MR caused by the Kondo effect can be expressed as
\begin{equation} \label{yosida}
    \Delta\rho/\rho=-S(T)B^2
\end{equation}

where $S(T)$ is a temperature-dependent coefficient given by~\cite{Yosida,YKatayama_PRL_1966,YKatayama_PR_1967},

\begin{equation}\label{ST}
    S(T)^{-1/2}\propto(T+T_0)
\end{equation}

where $T_0$ is a positive parameter. We obtained $S(T)$ by fitting MR with equation~\ref{yosida} at different temperatures below $T_m$. Figure \ref{fig2}c) shows $ S(T)^{-1/2}$ as a function of $T$. This data shows reasonable agreement with linear fitting, demonstrating good agreement with Yosida's theory. From this analysis, we conclude that, at low fields $\leq4$ T the magnetoresistance in Yb\textsubscript{0.5}Co\textsubscript{3}Ge\textsubscript{3} results from Kondo scattering. However, several questions still remain regarding the origin of the Kondo scatterers in this sample.

Usually, in a metallic diffusive system, Kondo scattering is attributed to spin-flip scattering by magnetic impurities. In this case, there are no known magnetic impurities. Rather, two types of magnetic atoms, Yb and Co periodically appear in the crystal, and one of these two moments is probably responsible for the Kondo effect. It was indeed previously confirmed experimentally that at least at low fields ($0.1$~T), Yb\textsuperscript{3+} moments contribute to magnetic susceptibility, and they interact with each other through antiferromagnetic exchange interaction with \textit{c}-axis as the easy axis~\cite{JChan}. The magnetoresistance in our device also shows an anisotropic behavior with stronger MR signature along the \textit{c}-axis (Figure~\ref{fig:Ext_MR_angle}), in agreement with previous measurements~\cite{Wang_Kagome_Yb}. Along with this, the field scale up to which the MR shows a parabolic fitting ($\sim4$ T) also matches well with the onset of saturation previously found in the magnetic field dependence of susceptibility along \textit{c}-axis~\cite{JChan}. When all of the above are taken into account, it becomes clear that the low-field Kondo effect is caused by Yb\textsuperscript{3+} moments. However, Yb is not an impurity in the crystal. Rather, The periodic arrangement of Yb can lead to a tantalizing possibility--the formation of a Kondo lattice system~\cite{coleman1509heavy_Kondo_lattice}. It is well-understood that when a Kondo lattice is cooled down so that phonon scattering is suppressed, the incoherent spin-flip scattering by antiferromagnetically-coupled magnetic moment would lead to Kondo effect with a $\sim ln(T)$ type behavior as observed by us. In a dense Kondo lattice system, at even lower temperatures, the intersite-coupling between these moments, mediated by the Kondo screening cloud around them, leads to the formation of a lattice coherent state and will result in a downturn of resistivity~\cite{jang2020evolution}. So far in our measurements, we do not observe any evidence of this coherent state, and this can be attributed to either not accessing low enough temperatures, or high enough density. We decided to investigate whether the exchange interaction between these moments can be enhanced by pressurizing the sample. 

The lack of saturation of MR at high fields up to 8 T is not well-understood. However, we note that the previous field-dependence of susceptibility also does not completely saturate up to 7 T even when the field is applied along the easy axis~\cite{JChan}. Further measurements, beyond the scope of this paper, are required to understand this unsaturating behavior.

Next, we carried out pressure-dependent electrical transport measurements in a piston-cylinder pressure cell Figure \ref{fig3}. Figure \ref{fig3}a) shows $\rho$ - $T$ at different pressures in the range from 0 GPa to 2 GPa. We note two interesting observations. First, $T_m$, the onset temperature of resistance upturn increases with increasing pressure (Figure\ref{fig3}b). Second, with increasing pressure, a downturn of resistance starts appearing at $T^*\sim4$~K starting at 1 GPa. Both of these phenomena can be attributed to the enhancement of the coupling between Yb\textsuperscript{3+} moment and the Fermi surface due to pressure. This leads to an enhancement of spin-flip scattering and hence a rise of $T_m$ at the higher temperature range. At lower temperatures ($T^*$), this leads to an enhancement of exchange interaction and results in an onset of coherent lattice state, as reported previously in Ce\textsubscript{2}Pd\textsubscript{3}Si\textsubscript{5}~\cite{MAbliz2006}. It is important to note that the magnitute of the downturn does not change much between 1.5 GPa and 2 GPa. This can be attributed to the solidification of the hydrostatic pressure medium (daphne oil 7373) at $~2$ GPa at room temperature~\cite{yokogawa2007solidification_Daphne7373}.

Figure \ref{fig3}c) shows magnetoresistance measurement as a function of pressure at 2 K. The enhancement of the magnetoresistance is consistent with the enhancement of the Kondo-exchange. 

To conclude, we have performed magnetotransport measurements on Yb\textsubscript{0.5}Co\textsubscript{3}Ge\textsubscript{3} Kagome metal samples. Our measurements indicate a low temperature Kondo effect in this system which most likely transitions to a Kondo coherence phase upon application of pressure.

\newpage
\bibliography{References}

\begin{thebibliography}{10}

\bibitem{Yin_Kagome_rev}
Jia-Xin Yin, Biao Lian, and M~Zahid Hasan.
\newblock Topological {K}agome {M}agnets and {S}uperconductors.
\newblock {\em Nature}, 612(7941):647--657, 2022.

\bibitem{Wang_Kagome_rev}
Yaojia Wang, Heng Wu, Gregory~T McCandless, Julia~Y Chan, and Mazhar~N Ali.
\newblock Quantum {S}tates and {I}ntertwining {P}hases in {K}agome {M}aterials.
\newblock {\em Nature Reviews Physics}, 5(11):635--658, 2023.

\bibitem{zhao2021cascade_Kagome_dope_1}
He~Zhao, Hong Li, Brenden~R Ortiz, Samuel~ML Teicher, Takamori Park, Mengxing Ye, Ziqiang Wang, Leon Balents, Stephen~D Wilson, and Ilija Zeljkovic.
\newblock Cascade of {C}orrelated {E}lectron {S}tates in the {K}agome {S}uperconductor {C}s{V}$_{3}${S}b$_{5}$.
\newblock {\em Nature}, 599(7884):216--221, 2021.

\bibitem{li2023unidirectional_Kagome_dope_2}
Hong Li, He~Zhao, Brenden~R Ortiz, Yuzki Oey, Ziqiang Wang, Stephen~D Wilson, and Ilija Zeljkovic.
\newblock Unidirectional {C}oherent {Q}uasiparticles in the {H}igh-{T}emperature {T}otational {S}ymmetry {B}roken {P}hase of {A}{V}$_{3}${S}b$_{5}$ {K}agome {S}uperconductors.
\newblock {\em Nature Physics}, 19(5):637--643, 2023.

\bibitem{li2022discovery_Kagome_pressure_1}
Haoxiang Li, Gilberto Fabbris, AH~Said, JP~Sun, Yu-Xiao Jiang, J-X Yin, Yun-Yi Pai, Sangmoon Yoon, Andrew~R Lupini, CS~Nelson, et~al.
\newblock Discovery of {C}onjoined {C}harge {D}ensity {W}aves in the {K}agome {S}uperconductor {C}s{V}$_{3}${S}b$_{5}$.
\newblock {\em Nature communications}, 13(1):6348, 2022.

\bibitem{zheng2022emergent_Kagome_pressure_2}
Lixuan Zheng, Zhimian Wu, Ye~Yang, Linpeng Nie, Min Shan, Kuanglv Sun, Dianwu Song, Fanghang Yu, Jian Li, Dan Zhao, et~al.
\newblock Emergent {C}harge {O}rder in {P}ressurized {K}agome {S}uperconductor {C}s{V}$_{3}${S}b$_{5}$.
\newblock {\em Nature}, 611(7937):682--687, 2022.

\bibitem{JChan}
Ashley Weiland, Lucas~J. Eddy, Gregory~T. McCandless, Halyna Hodovanets, Johnpierre Paglione, and Julia~Y Chan.
\newblock Refine {I}ntervention: {C}haracterizing {D}isordered {Y}b$_{0.5}${C}o$_{3}${G}e$_{3}$.
\newblock {\em Crystal Growth \& Design}, 20(10):6715--6721, 2020.

\bibitem{tan2024three}
Hengxin Tan, Yiyang Jiang, Gregory~T McCandless, Julia~Y Chan, and Binghai Yan.
\newblock Three-dimensional higher-order saddle points induced flat bands in co-based kagome metals.
\newblock {\em arXiv preprint arXiv:2405.04863}, 2024.

\bibitem{Wang_Kagome_Yb}
Yaojia Wang, Gregory~T. McCandless, Xiaoping Wang, Kulatheepan Thanabalasingam, Heng Wu, Damian Bouwmeester, Herre S.~J. van~der Zant, Mazhar~N. Ali, and Julia~Y. Chan.
\newblock Electronic {P}roperties and {P}hase {T}ransition in the {K}agome {M}etal {Y}b$_{0.5}${C}o$_{3}${G}e$_{3}$.
\newblock {\em Chemistry of Materials}, 34(16):7337--7343, 2022.

\bibitem{PLee&TRamakrishnan}
Patrick~A. Lee and T.~V. Ramakrishnan.
\newblock Disordered {E}lectronic {S}ystems.
\newblock {\em Rev. Mod. Phys.}, 57:287--337, 1985.

\bibitem{YKatayama_PRL_1966}
Yoshifumi Katayama and Shoji Tanaka.
\newblock Resistance {A}nomaly in $n$-{T}ype {I}n{S}b at {V}ery {L}ow {T}emperatures.
\newblock {\em Phys. Rev. Lett.}, 16:129--131, 1966.

\bibitem{YKatayama_PR_1967}
Yoshifumi Katayama and Shoji Tanaka.
\newblock Resistance {A}nomaly and {N}egative {M}agnetoresistance in $n$-{T}ype {I}n{S}b at {V}ery {L}ow {T}emperatures.
\newblock {\em Phys. Rev.}, 153:873--882, 1967.

\bibitem{mani1991weak_3D_WL}
RG~Mani, L~Ghenim, and JB~Choi.
\newblock Weak {L}ocalization in the {N}arrow {G}ap {B}ulk {S}emiconductors {H}g$_{1-x}${C}d$_{x}${T}e and {I}n{S}b.
\newblock {\em Solid state communications}, 79(8):693--697, 1991.

\bibitem{Yosida}
Kei Yosida.
\newblock Anomalous {E}lectrical {R}esistivity and {M}agnetoresistance {D}ue to an $s\ensuremath{-}d$ {I}nteraction in {C}u-{M}n {A}lloys.
\newblock {\em Phys. Rev.}, 107:396--403, 1957.

\bibitem{coleman1509heavy_Kondo_lattice}
P~Coleman.
\newblock Heavy {F}ermions and the {K}ondo {L}attice: a 21st {C}entury {P}erspective (2015).
\newblock {\em arXiv preprint arXiv:1509.05769}.

\bibitem{jang2020evolution}
Sooyoung Jang, JD~Denlinger, JW~Allen, Vivien~S Zapf, MB~Maple, Jae~Nyeong Kim, Bo~Gyu Jang, and Ji~Hoon Shim.
\newblock Evolution of the kondo lattice electronic structure above the transport coherence temperature.
\newblock {\em Proceedings of the National Academy of Sciences}, 117(38):23467--23476, 2020.

\bibitem{MAbliz2006}
Melike Abliz, Masato Hedo, Jiro Kitagawa, Yoshiya Uwatoko, and Masayasu Ishikawa.
\newblock Pressure {I}nduced {K}ondo {C}oherence {E}ffect in {C}e$_{2}${P}d$_{3}${S}i$_{5}$.
\newblock {\em Journal of Alloys and Compounds}, 408-412:241--243, 2006.

\bibitem{yokogawa2007solidification_Daphne7373}
Keiichi Yokogawa, Keizo Murata, Harukazu Yoshino, and Shoji Aoyama.
\newblock Solidification of {H}igh-{P}ressure {M}edium {D}aphne 7373.
\newblock {\em Japanese journal of applied physics}, 46(6R):3636, 2007.

\end{thebibliography}
\bibliographystyle{unsrt}

\newpage
\setcounter{figure}{0}
\renewcommand{\figurename}{Fig.}
\renewcommand{\thefigure}{E\arabic{figure}}

\begin{figure*}[h]
    \centering
    \includegraphics[width=\textwidth]{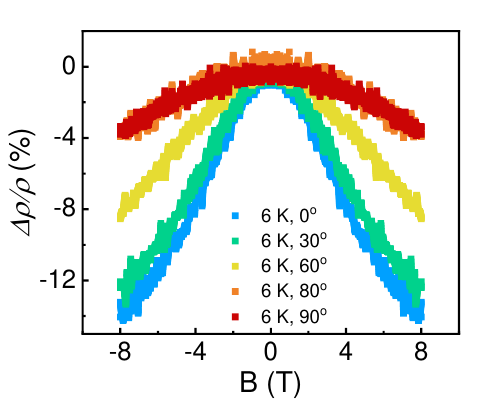}
    \caption{Angle-dependent MR measurements}
    \label{fig:Ext_MR_angle}
\end{figure*}

\end{document}